\documentclass[twocolumn]{article}

\pdfoutput=1

\usepackage{amsmath}
\usepackage{amssymb}
\usepackage{graphicx}
\usepackage{hyperref}

\newcommand{\norm}[1]{\left\lVert#1\right\rVert}
\DeclareMathOperator*{\argmin}{arg\,min}

\date{}

\author{
  Demir, \"Ozg\"ur \\
  \small{\url{demir@soundcloud.com}}
  \and
  Rodriguez Yakushev, Alexey \\
  \small{\url{alexey.rodriguez.yakushev@soundcloud.com}}
  \and
  Keddo, Rany\\
  \small{\url{rany@soundcloud.com}}
  \and
  Kallio, Ursula \\
  \small{\url{ursula@soundcloud.com}}
}

\title{Item-Item Music Recommendations With Side Information}

\begin{document}

\maketitle
\pagenumbering{gobble}
\newpage
\pagenumbering{arabic}

\begin{abstract}
Online music services have tens of millions of tracks. The content itself is broad and covers various musical genres as well as non-musical audio content such as radio plays and podcasts. The sheer scale and diversity of content makes it difficult for a user to find relevant tracks. Relevant recommendations are therefore crucial for a good user experience.
Here we present a method to compute track-track similarities using collaborative filtering signals with side information.
On a data set from music streaming service SoundCloud, the method here outperforms the widely adopted implicit matrix factorization technique.
The implementation of our method is open sourced and can be applied to related item-item recommendation tasks with side information.
\end{abstract}


\section{Introduction}
Detecting music similarity can be divided into three main approaches: content-based, metadata-based, and collaborative filtering.

A content-based approach utilizes raw audio material to detect track-track similarities \cite{content-based}. However, current methods are only able to distinguish top level genres from each other rather than sub-categories, and therefore the content-based approach does not lead to fine-grained recommendations \cite{ismir-2010}.

A metadata-based approach relies on the metadata that is associated with the audio signal, such as creator, year of release, genre, and tempo. The metadata tends to be very sparse. Additionally, a tag can be incorrect. As a result of these two points, any metadata-based approach on its own does not lead to relevant recommendations.

A collaborative filtering approach incorporates user-behavior data, and it is the most widely used approach among recommender systems for audio data because it often outperforms the other two approaches \cite{journals/ieeemm/Slaney11}. However, this approach suffers from the cold-start problem: too few users have interacted with a given track, which makes it difficult to relate it to other tracks \cite{cold-start}. That said the majority of user interactions apply to a small percentage of the catalog \cite{long-tail}. Therefore, on its own collaborative filtering already provides accurate results for the majority of user interactions. Combining collaborative filtering with content- and metadata-based approaches may still improve recommendations because the content- and metadata-based approaches do not suffer from the cold-start problem.

Most music recommender systems are personalized. Based on a user's past activity, they recommend tracks a user might find interesting. For new and anonymous users, previous activity is unknown. Therefore, many streaming services display non-personalized recommendations, such as the most similar track to the one currently playing. For these reasons, non-personalized, track-track recommendations play an important role.

A simple approach might be to count how frequently users have listened to track $b$ after having listened to track $a$. Next, normalize the count by the overall popularity of both tracks \cite{Davidson:2010:YVR:1864708.1864770, Spertus:2005:ESM:1081870.1081956}.
The main disadvantage of this approach is that it is not possible to relate two tracks to each other without a significant amount of co-interactions e.g. two items might be very similar to each other through an intermediary item but have never been interacted with together.

Latent factor models try to overcome this problem by finding an embedding that is a vector representation for each item in the train set. These vectors are ordered such that items considered similar by the community are nearby in the high dimensional vector space. A popular latent factor algorithm for implicit feedback signals is described in \cite{implicit-factorization}. The objective of this model is as follows:

\begin{equation}
 \argmin_{\vec{u}, \vec{p}} \sum_{r_{ij}} c_{ij}(r_{ij} - \vec{u}_i^t \vec{p}_j)^2 + \lambda_1 \norm{\vec{u}}^2 + \lambda_2  \norm{\vec{p}}^2,
\end{equation}
where $r_{ij} = 1$ if user $u_i$ interacted with product $p_j$, and $r_{ij} = 0$ otherwise. $c_{ij}$ defines the confidence of user $u_i$ for product $p_j$. $c_{ij}$ increases if the user interacted with an item more often. This model results in user- and item-embeddings where the latter can be used to detect item-item similarities.

\section{Methods}

\subsection{User Interaction Model} \label{user-interaction-model}

Most music services have multiple sources of user-feedback signals, which can be categorized as implicit (plays, shares, comments, playlist additions, page visits, search queries etc.) or explicit (likes and ratings). It is not obvious how to combine these signals. One user might find it important to share a track, whereas another user might find it important to like it. 

To overcome this, we propose a simple user-interaction model. In this model, all of the actions that a user performed on an particular item are merged. The resulting merged event is positive, if it is a strong positive interaction such as a playlist addition, a share, or a like. It is also a positive event if the user fully listened to the same track more than once. A single listen might have occurred unintentionally (such as letting a song play while not paying attention) and therefore does not indicate a strong positive signal. A full listen is defined as such when the user listened to either the majority of the track or for a long absolute time span, such as 20 minutes of the duration of a podcast. This user-interaction model only keeps strong positive signals, and noise is filtered out at an early stage. The downside is that this model will create some false negatives which can be compensated by the huge amount of available user-interactions. 

It might seem obvious to use skips as negative signals. However, we found that skips strongly correlate with positive interactions. A possible explanation is that users  skip items with which they interact frequently. For example, a user chooses to skip to a specific part of a track, such as to its refrain.

The positive user-interactions are further filtered: users with fewer than 5 interacted items, and items with fewer than 5 interactions from unique users are removed. Collaborative filtering is based on co-occurrences; users who have too few interactions will not contribute any meaningful signal, and items with too few interactions will not get relevant recommendations. Item-interactions are further filtered by randomly sampling 10,000 interactions per item. This is mainly to speed up computation time because some popular items might have up to millions of interactions associated with them. Moreover, this sampling has the benefit of avoiding that a handful of tracks overly dominate the loss function during training. The additional filtering steps dramatically reduce the size of the train set and further reduce the effect of noise. 

This user-interaction model produces a set of positive, user-track interaction pairs.

\subsection{Train Set} \label{train-set}
User consumption patterns are domain specific. For example, when a user watches a movie they are less likely to watch it again immediately. On the other hand, a user commonly listens to a song repeatedly. Furthermore, a user prefers a listening experience with smooth transitions, rather than one that alternates between songs of vastly different characteristics such as genre, tempo, or mood. However, over time a user might gradually transition from one genre to another that is vastly different. The same can be said about any musical characteristic.

\begin{figure}[t]
  \centering
  \includegraphics[scale=0.8]{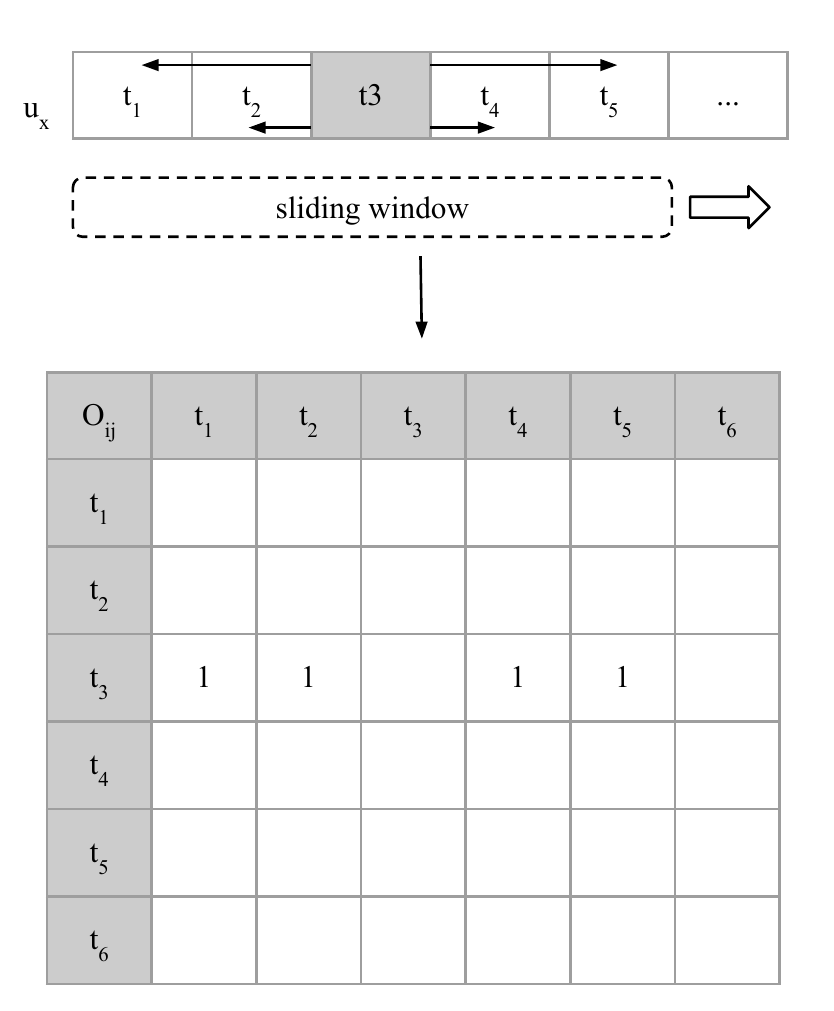}
  \caption{Co-occurrence matrix $O$. A sliding window is moved along positive user ($u_x$) track ($t_y$) interactions. Pairs are generated from the central track and each of its surrounding tracks and the pair occurrence is stored in a track-track interaction matrix $O_{ij}$.}
  \label{fig:sliding-window}
\end{figure}

In order to capture the consumption pattern in which track similarity is stronger over short periods of time but less so over longer periods, we use a sliding window that restricts the way that training examples are sampled. This sampling technique is directly inspired by work in the word embeddings community \cite{word2vec, glove, swivel}. The sliding window moves along the positive user-track interactions, which are ordered by interaction start time. The window size can be track- or time-based. With each discrete window slide, pairs are generated from the central track and each of its surrounding tracks. Pair occurrences are accounted for in a track-track co-occurrence matrix $O$, whose entries $O_{ij}$ denote the number of times track $i$ has been listened to in the context of track $j$ (figure \ref{fig:sliding-window}). Pairs emerging from more distant window positions are less related to each other and may be down weighted while counting. We define $O^+ \subseteq O$ with $O^+_{ij} > 0, \forall i,j$ as the set of all observed pairs.

\subsection{Latent Factor Model}

\begin{figure}[t]
  \centering
  \includegraphics[scale=0.8]{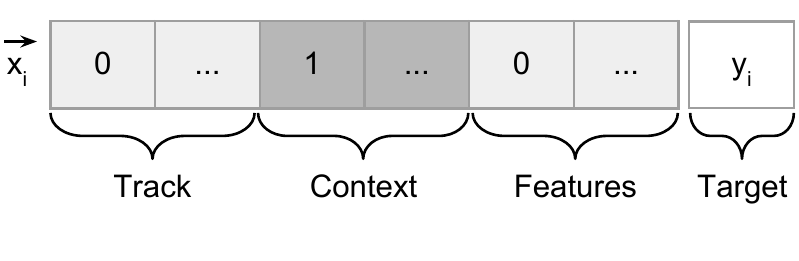}
  \caption{Example training instance. Each of the three parts represents a one-hot encoding: the first part encodes the track, the second part encodes the context track and the third part encodes the track creator. $y_i$ is the target value.}
  \label{fig:instance-vector}
\end{figure}

The co-occurrence matrix $O^+$ is used in a latent factor model. Our implementation is based on factorization machines \cite{factorization-machines}. The prediction function is defined as follows:

\begin{equation} \label{eq:fm}
	\hat{y}(\vec{x})= \sum_{i=1}^n w_i x_i +
	\sum_{i=1}^n\sum_{j=i+1}^n \vec{v_i}^\intercal \vec{v_j} x_i x_j.
\end{equation}
\begin{itemize}
  \item $\vec{x}$ is a vector of binary indicator variables that are generated for each entry in $O^+$ and is encoded thusly (figure \ref{fig:instance-vector}): 
  \begin{itemize}
	  \item The first part of $\vec{x}$ encodes a track.
	  \item The second part encodes a context track.
	  \item The third part encodes additional features as described below.
	  \end{itemize}
  \item $w_i$ can be regarded as a bias term.
  \item  $\vec{v}_i$ is a latent track vector, and $\vec{v}_j$ is a latent context track vector. 
   
   $|\vec{v}| = k \in \mathbb{N}^+$, where $k$ is a user-defined parameter.
\end{itemize}
In standard matrix factorization, the feature vector only includes the user and the track. Hence, there is only a single non-zero interaction term per training instance. In factorization machines, the objective function can handle an arbitrary number of features in the feature vector. The additional features produce an additional number of non-zero interaction terms. Additional features that are used in the function can be meta data based or content based. Meta data based information can include the track's length, its age, its popularity, or its creator. Content-based information can include the track's audio features such as mel frequency cepstral coefficients, octave based spectral contrast coefficients or chroma features \cite{content-based}.
In equation \ref{eq:fm}, each feature results in an additional latent vector for a particular track; $|\vec{x}| = n = 2C+F$, where $C$ is the catalog size, and $F$ is the number of additional features per track. The final vector of a track is the sum of all of the track's latent vectors. The training objective is defined as follows:

\begin{equation} \label{eq:obj}
 \begin{split}
\argmin_{\vec{x}}
	E(\vec{x})= \sum_{i=1}^N p_i L(\hat{y}(\vec{x}_i), y_i)) + \\
	\lambda_1 \norm{\vec{w}}^2_2 +
	\lambda_2 \sum_{i=1}^n \norm{\vec{v}_i}^2_2,
\end{split}
\end{equation}
where $L(\hat{y},y)$ is a loss function that measures the discrepancy between the predicted value $\hat{y}$ and the target value $y$. $L$ can be a regression loss, such as a squared loss: $L(\hat{y},y) = (\hat{y} - y)^2$. When using a regression loss, the target value can be set to $log_2(O^+_{ij})$ \cite{glove}. This training objective places positive pairs near each other. However, it does not guarantee that unrelated pairs are far apart \cite{swivel}. 

A classification loss, such as the logistic loss, can be used to place unrelated pairs far apart from each other: $L(\hat{y},y) = log(1 + exp(-\hat{y}y))$. Negative interactions are thereby sampled from $O^+$, which is performed non-uniformly according to a smoothed track-occurrence distribution \cite{word2vec}. Due to the large size of the catalog, two randomly sampled items are expected to be dissimilar. When using a classification loss the target value $y$ can be defined as follows:

\begin{equation}
	y_i = \left\{
		\begin{array}{l l}
			1 & \quad \mbox{if $\vec{x}_i$ is observed}\\
			-1 & \quad \mbox{if $\vec{x}_i$ is sampled.}\\
		\end{array}
		\right.
\end{equation}

The objective function is minimized by means of a stochastic gradient decent approach. In each step, the learning rates are adjusted according to the AdaGrad scheme \cite{adagrad}. However, in order to save computation time and memory consumption, the same learning rate is applied to all dimensions of a vector. For a matrix factorization task, this approach is as effective as assigning a learning rate for each vector dimension \cite{mf-adagrad}.

\section{Experiments}

The proposed approach was trained on a proprietary real-world data set that was extracted from data logs of the music streaming service SoundCloud (\url{http://soundcloud.com}). The data set contains approximately 1.5 billion interactions from about 40 million users. The data set has been pre-processed as described in section \ref{user-interaction-model} in order to extract pairs of positive user-track interactions. We split the log data at a specific timestamp; all of the interactions before the split point are used to train models, and all of the interactions after the split point are used to test these models. The test set contains 5 million users with a total of 25 million interactions. The tracks that a user has interacted with in the train set have been removed from the test set.

The item-item model was built using a logistic loss with five negative samples. The sliding window size was set to 10 items (five to the left, five to the right). This model is referred to here as \textit{ITEM}.
Another item-item model was trained where the creator of the track was added as additional side information. This model is referred to here as \textit{ITEMc}, where c denotes creator.

As a baseline method, we trained the implicit matrix factorization method defined in \cite{implicit-factorization}. The $\alpha$ parameter was set to balance out the positive and negative interactions. This model is referred to here as \textit{IMPL}.

The regularization parameter $\lambda$ of each model was tuned based on an independent validation set that was extracted from the test set. The number of latent factors was set to 150 for each model.

The mean percentile rank (MPR) metric was used to evaluate models \cite{implicit-factorization}. The percentile rank of a track $a$ is computed as follows:

\begin{equation}
	pr_a = \frac{\sum_{b,i} H(cos(\vec{x}_a,\vec{x}_i) - cos(\vec{x}_a,\vec{x}_b))}{\sum_{b,i} 1},
\end{equation}
where $b$ are tracks which have been played in the context of track $a$, and $i$ are all remaining tracks. $cos(\vec{x}, \vec{y})$ is the cosine similarity between $\vec{x}$ and $\vec{y}$. $H(x)$ is the Heaviside step function. As previously mentioned, tracks that are listened to in the same context are deemed similar to each other. The MPR is a measurement of how close similarly deemed tracks are situated in the latent model space.
The average MPR comprises predominantly popular tracks. However, most models do not provide adequate results in the long tail, where there are only a few interactions. Therefore, we computed MPR values of various bins. A track is assigned to a bin based on its total number of occurrences in the train set.

\section{Results}

Table \ref{table:mpr} shows the MPR of all three models. The proposed item-item model (\textit{ITEM}) clearly outperforms the standard implicit feedback model (\textit{IMPL}) for all bins. Overall, the performance drops for tracks with fewer signals. Adding side information improves the results furthermore (\textit{ITEMc}). This is especially the case for the long tail where fewer user-interaction signals are present. Tracks that occur only five times in the train set have significantly better recommendations than by showing tracks at random (MPR = 0.5).

\begin{table}[t]
  \begin{tabular}{ll|ccc}
    bin$^1$ & count$^2$ & \textit{IMPL} & \textit{ITEM} & \textit{ITEMc}$^3$ \\
    \hline
5     & 3583   & 0.1680 & 0.0970 & \textbf{0.0865} \\
10    & 14670  & 0.1487 & 0.0818 & \textbf{0.0756} \\
20    & 22009  & 0.1282 & 0.0669 & \textbf{0.0580} \\
50    & 42156  & 0.1125 & 0.0568 & \textbf{0.0509} \\
100   & 43188  & 0.1012 & 0.0481 & \textbf{0.0449} \\
1000  & 225754 & 0.0949 & 0.0433 & \textbf{0.0410} \\
5000  & 193858 & 0.0803 & 0.0428 & \textbf{0.0400} \\
15000 & 337153 & 0.0828 & 0.0419 & \textbf{0.0402} \\
\hline
avg. & 884411 & 0.0901 & 0.0449 & \textbf{0.0422} \\
  \end{tabular}
  \caption{MPR of the three models. $^1$ A track is assigned to a bin based on its total number of occurrences in train set. $^2$ The total number of tracks per bin. $^3$ The best (lowest) values per bin are in bold.}
  \label{table:mpr}
\end{table}

\section{Conclusions}
Personalized latent factor models have been widely adopted for the music-recommendations domain. These models compute latent vector embeddings for  both users and items. While the later ones can be used to compute item-item similarities, we showed that a model learned on the item-item co-occurrence matrix yields better performance. This matrix is computed by considering domain consumption patterns via a sliding window, and this matrix is less sparse than the user-item interactions matrix. Available side information, which on its own might not be very accurate, increases the recommendation quality, especially in the long tail.

Though the presented method generates a non-personalized item-item model, a simple yet effective way of generating personalized recommendations is to  recommend tracks that are similar to tracks with which the user previously interacted. This way, an explanation such as \textit{"since you liked track x you also might like track y"} can be displayed to the user.

The implementation of our method is open sourced and can be applied to related item-item recommendation tasks with side information (\url{https://github.com/ozgurdemir/item-item-factorization}).

\section*{Acknowledgement}
We thank Christoph Sawade for proof reading and fruitful discussions.

\bibliography{item-item-factorization}
\bibliographystyle{plain}
\end{document}